# A Deep Learning Model of Lightning Stroke Density

Randall Jones II[1], Joel A. Thornton[1], Chris J. Wright[1], Robert A. Holzworth[2]

[1]University of Washington, Department of Atmospheric and Climate Sciences, [2]University of Washington, Department of Earth and Space Sciences

## ABSTRACT

Lightning plays a crucial role in the Earth's climate system, yet existing parameterizations for use in forecasting and earth system models show room for improvement in capturing spatial and temporal variations in its frequency. This study develops deep learning-based parameterizations of lightning stroke density using meteorological variables from the ERA and IMERG datasets. Convolutional neural networks (CNNs) with U-Net architectures are trained using World Wide Lightning Location Network (WWLLN) data from 2010 to 2021 and evaluated on WWLLN lightning observations from 2022 and 2023. The CNNs reduce the average domain mean bias by an order of magnitude and produce significantly higher Fractions Skill Score (FSS) values across all lightning regimes compared to the multiplicative product of CAPE and precipitation. The CNNs show skill relative to previously published parameterizations over the oceans especially, with $r^2$ values as high as 0.93 achieved between the best performing modeled and observed lightning stroke density climatologies. The CNNs are also able to accurately capture the 12-hourly evolution of lightning spatial patterns on an event-by-event basis with high skill. These results show the potential for deep learning to improve on lightning parameterizations in weather and earth system models.

## INTRODUCTION

Lightning is one of the most powerful and destructive phenomena that frequently occurs on Earth and its observation is used as an indicator of severe weather. Annual fatality rates as high as 84 deaths per million per year have been reported in lesser developed nations in recent times [*Borgerhoff Mulder et al.*, 2012; *Holle*, 2016]. Lightning is also estimated to be responsible for 56% of total burned area in the continental United States between the years 1992 and 2012, despite only being the cause for 16% of these wildfires [*Balch et al.*, 2017].

Besides direct impacts on human safety, lightning plays a key role in the climate system. Lightning strokes are a globally important direct source of reactive trace gases such as nitrogen oxide radicals, hydroxyl radicals, and ozone which regulate the self-cleansing power of the troposphere [*Logan,* 1983; *Mickley et al.,* 2001; *Yuan et al.,* 2012; *Stevenson et al.*, 2013; *Mao et al.*, 2021]. The oxidizing power of these trace gases regulates the formation of tropospheric ozone and the lifetime of methane, key anthropogenic greenhouse gases [*Schumann and Huntrieser*, 2007, *Yuan et al.,* 2012]. In addition, this chemistry has the potential to impact the formation and growth of cloud condensation nuclei (CCN) in deep convective outflow, which is now considered an important global source of CCN [*Phillips et al.*, 2007; *Bardakov et al.*, 2024], which in turn affect Earth's energy balance directly and indirectly through changes to cloud optical properties and lifetimes.

Understanding the environmental drivers of lightning is therefore important in a changing climate. Recent studies found different factors are more strongly related to lightning over land than over ocean, as well as for specialized domains [*Romps et al.,* 2014; *Stolz et al.,* 2015; *Stolz et al.,* 2017; *Thornton et al.,* 2017; *Romps et al.,* 2018; *Etten-Bohm et al.*, 2021; *Cheng et al.,* 2021]. Romps et al. (2014) found that the multiplicative product of convective available potential energy (CAPE), precipitation rate and a constant of proportionality was able to explain 77% of the variance at a 12-hourly temporal resolution in cloud-to-ground lightning flash rate in the continental United States (CONUS), obtained from the National Lightning Detection Network, for the year 2011. A linear regression of lightning stroke density to the multiplicative product of CAPE and precipitation will hereafter be referred to as the R14 parameterization. However, Romps et al. (2018) shows that this parameterization does not work particularly well over the ocean, overestimating lightning in the tropical oceans, pointing to the possibility of other variables being responsible for this variance.

Other researchers have aimed to improve lightning prediction through the inclusion of additional variables. Cheng et al. (2021) implemented a CAPE threshold over the tropical oceans, which was able to more accurately capture the land-sea contrast in lightning stroke density than the Romps et al. (2014) study, but possibly points to other variables being influential, such as aerosol particle concentrations and size distributions, which have been linked to temporal and spatial variations in lightning stroke density in other studies [*Stolz et al.,* 2015; *Proestakis et al.,* 2016; *Stolz et al.,* 2017; *Altaratz et al.,* 2017; *Thornton et al.,* 2017; *Pan et al.,* 2022; *Wright et al.*, 2025].

Stolz et al. (2015) investigated the sensitivity of total lightning density from the Tropical Rainfall Measuring Mission's (TRMM) lightning imaging sensor (LIS) to normalized CAPE, simulated lower tropospheric aerosol number concentrations with diameters larger than 40 nanometers from GEOS-Chem and warm cloud depth, defined as the difference between the lifting condensation level and the freezing level, globally between 36°S and 36°N between the years 2004 and 2011. They found that the highest total lightning density occurred in areas with above average normalized CAPE, heightened concentrations of aerosols with diameters larger than 40 nanometers, and shallower warm cloud depth.

Stolz et al. (2017) expanded upon this study, investigating the relative contributions of normalized CAPE, CCN concentrations, warm cloud depth, vertical wind shear and environmental relative humidity to the variability of total lightning density and average 30dbZ radar reflectivity from the TRMM LIS. The results of multiple linear regression methods show that lightning density and average 30dbZ radar reflectivity increase with increasing normalized CAPE, CNN and vertical wind shear, and decreasing warm cloud depth and relative environmental humidity, with $r^2$ values reaching 0.81.

Similar to the work done in Stolz et al. (2017), Etten-Bohm et al. (2021) aimed to create a lightning parameterization using lightning from TRMM LIS and CAPE, normalized CAPE, lifting condensation level, column saturation fraction, low-level wind shear, deep wind shear and 700-hPa omega using a logistic regression technique from the 3-hourly Modern-Era Retrospective analysis for Research and Application version 2 (MERRA-2) between 35°S and 35°N from 1998 to 2013. They found that an increase in lightning was linked with an increase in all environmental variables with the exception of shear. Additionally, with the inclusion of geographic variables, such as coast and terrain, the logistic regression method accurately predicted lightning occurrence 86% of the time.

The goals of our study are similar to those of Stolz et al. (2017) and Etten-Bohm et al. approaches, but we further assess whether straight-forward deep learning algorithms add significant value in the development of lightning parameterizations. We use similar variable sets except for CCN concentrations, which we exclude. The goal of this study is to develop a convolutional neural network (CNN) machine learning algorithm to accurately capture the distribution and magnitude of lightning features over a relatively large domain, high spatial resolution, and 12-hourly timescales.

CNNs have increasingly been used in meteorology for a number of purposes, ranging from radar-based precipitation nowcasting to tropical cyclone intensity estimation [*Ayzel et al.,* 2010; *Jung et al.,* 2024]. A CNN is an image-based deep learning architecture that is particularly useful for pattern recognition. The specific CNN architecture we use in this study is a U-Net, which was introduced by Ronneberger et al. (2015) for biomedical image segmentation [*Ronneberger et al.,* 2015, *Weyn et al.,* 2020, *Chase et al.,* 2023]. U-Nets have since been widely adopted in the atmospheric sciences due to their ability to capture both local and global spatial patterns, making them well-suited for a range of Earth system processes which have recurring spatio-temporal patterns.

## METHODS
### Data

We use lightning data from the World Wide Lightning Location Network (WWLLN), a global network of very-low frequency radio lightning sensors operated by the University of Washington. WWLLN uses the time of group arrival of radio waves from at least five sensors to determine the location of lightning strokes, allowing for the location of a lightning strike to be accurate to 0.1°. Jacobson et al. (2006) showed that WWLLN reproduces roughly 70% of the spatial distribution of thunderstorms when aggregated over 3-hour periods, illustrating its usefulness for monitoring both intracloud and cloud-to-ground lightning on synoptic scales. Since 2006, WWLLN stroke detection efficiency has increased from near 1-3% globally to over 70% of all global strokes above 40 kA peak current and between 25-35% detection efficiency for all strokes, and the detection efficiency for all thunderstorms since 2009 has improved to well over 90% globally. Additionally, to address the nonuniform global detection efficiency of WWLLN, we apply the adjustment developed previously by Hutchins et al. (2012) which scales the observed stroke counts to represent the distribution expected from a uniformly sensitive global network.

We use CAPE, two-meter temperature (T2M), cloud base height (CBH), zero-degree level (ZDL), longitudinal and latitudinal wind components at the 500 and 1000 hPa pressure levels (U and V respectively), relative humidity (RH) at the 500 and 1000 hPa pressure levels and the land-sea mask (LSM) from the European Centre for Medium-Range Weather Forecasts (ECMWF)'s ERA5 Reanalysis. We derive warm cloud depth (WCD), defined as the difference between the ZDL and CBH, and wind shear (SHEAR), calculated using the formula $\sqrt{(u_{1000} - u_{500})^2 + (v_{1000} - v_{500})^2}$, from the ERA5 quantities and use these quantities in model runs as well. We use the average RH at the 500 and 1000 hPa pressure levels as the RH quantity for model runs.

We use precipitation data from the GPM IMERG Final Precipitation L3 Half Hourly 0.1° by 0.1° V07 product from NASA's Integrated Multi-SatellitE Retrievals for GPM (IMERG) algorithm, which provides hourly precipitation rates (mm hr$^{-1}$) at a 30-minute time resolution.

Our study domain ranges from 150° W to 0° longitude and 52° S to 52° N latitude, between the years 2010 and 2023. We chose this region for future comparison with the domain of the Geostationary Lightning Mapper (GLM) aboard NASA's GOES-18 satellite. Figure 1 shows the WWLLN lightning stroke density climatology for the focus region and time.

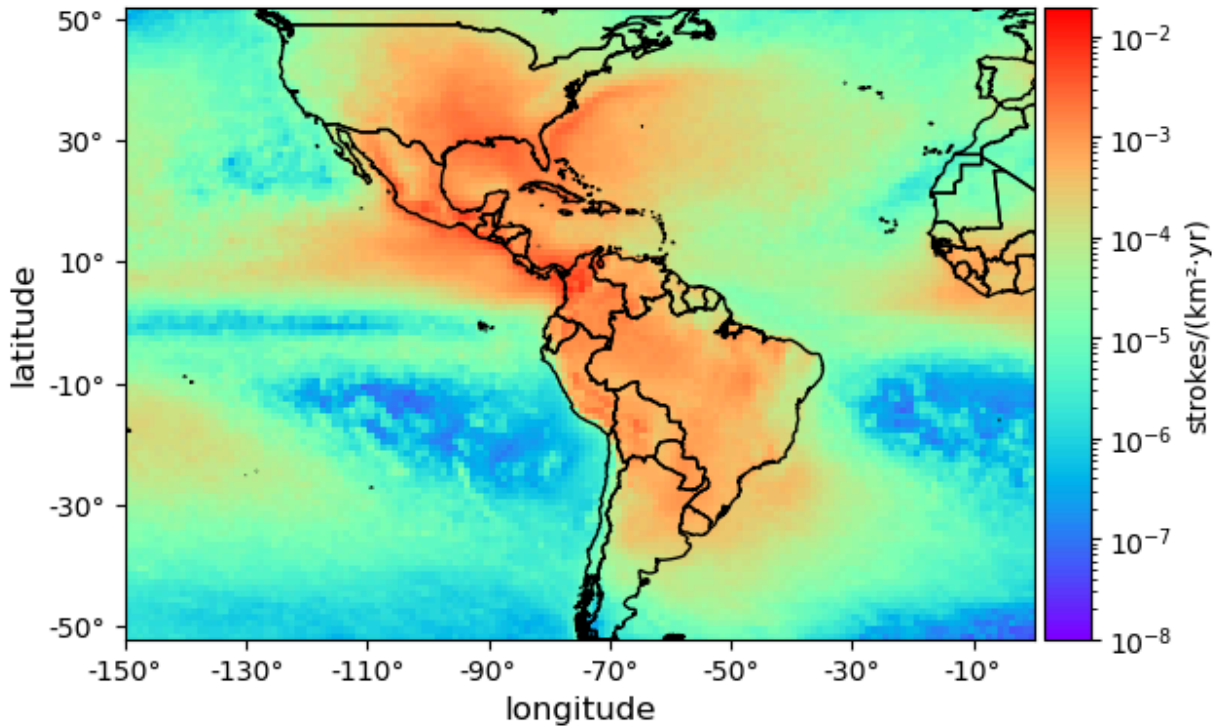

Figure 1: Mean WWLLN lightning stroke density (strokes km$^{-2}$ yr$^{-1}$) for the years 2010 to 2023 at a 0.5° latitude by 0.5° longitude spatial resolution.

We regridded all datasets using a first order conservative remapping to the desired spatial resolution. Table 1 summarizes dataset information.

| Dataset | Variable | Units | Native Spatial Resolution | Native Temporal Resolution |
|---|---|---|---|---|
| ERA5 | CAPE | J/kg | 0.25° x 0.25° | 1 hour |
| | T2M | K | | |
| | CBH | m | | |
| | ZDL | | | |
| | U | m/s | | |
| | V | | | |
| | RH | -- | | |
| | LSM | | | |
| IMERG | Precipitation | mm/hr | 0.1° x 0.1° | 30 minutes |
| WWLLN | Lightning | strokes km$^{-2}$ | | Instantaneous |
| | Detection Efficiency | N/A | 1° x 1° | Daily |

Table 1: Data sources we use in this study

**Fractional Skill Score**

To evaluate the performance of the model runs, we use the Fractions Skill Score (FSS), a spatial verification metric that quantifies the agreement between modeled and observed fields above a defined threshold. The FSS compares the fraction of grid points within a moving spatial window above a certain threshold in the modeled and observed fields [*Roberts and Lean,* 2008; *Roberts*, 2008; *Mittermaier et al.,* 2013]. The FSS is defined as:

$$FSS = 1 - \frac{\frac{1}{N}\sum_{i=1}^{N}(O_i - M_i)^2}{\frac{1}{N}\sum_{i=1}^{N}(O_i^2 + M_i^2)},$$

where $O_i$ and $M_i$ represent the observed and modeled fractions within the spatial window respectively, and N represents the number of windows in the domain. A value of 1 represents a perfect forecast and a value of 0 represents a forecast with no skill. In this study, we use a 3x3 window for all FSS calculations.

The FSS in its original form is used to calculate the fraction of observations above a certain threshold, but does not have an upper bound to show how close the predicted value is to the observed value. As such, we also use a modified FSS (MFSS) calculation that imposes an upper bound, which separates the data into bins based on orders of magnitude. For example, one bin will contain all points from $10^{-5}$ to $10^{-4}$ strokes km$^{-2}$ yr$^{-1}$ The FSS is then calculated separately for each bin, allowing us to assess model skill across different lightning stroke density ranges. This approach allows us to determine the skill of the models in the low and high lightning regimes.

**Machine Learning Model Configuration**

For the machine learning model runs, we reserve the years 2010 to 2021 for training and validation, while the years 2022 and 2023 are solely used for testing. Within the training and validation period, we randomly select 25% of the dataset for validation using a fixed seed, so each model run uses the same training and validation sets. Each variable used to train the CNN is given a corresponding letter for identification purposes: C for CAPE, P for precipitation, L for land-sea mask, R for 1000-500 hPa average relative humidity, S for 1000-500 hPa vertical wind shear, T for two-meter temperature, and W for warm cloud depth. For example, the CP CNN run would only contain CAPE and precipitation as input variables, and the CPL CNN run would contain CAPE, precipitation and the land-sea mask as input variables.

We normalize all input and output variables using z-score normalization in order to put all variables on a similar scale. We used z-score normalization in lieu of log-scaling the lightning data, which is another approach, due to log-scaled results producing a small number of large outliers which heavily skewed the output of the models and required imposing somewhat arbitrary removal of model output from the evaluations. Z-score normalization preserved the full dynamic range of the data while ensuring consistent scaling across variables.

We considered two different spatial resolutions in this study: 0.5° latitude by 0.5° longitude and 1° latitude by 1° longitude (hereafter, spatial resolution is reported as "latitudexlongitude°", where the numeric values represent degrees latitude by degrees longitude). We chose these spatial resolutions for computational efficiency. At these resolutions, model training takes up to two hours. However, we currently do not have the computing resources to run the model at a 3-hourly temporal resolution while also using a 0.5x0.5° spatial resolution, so three different combinations of temporal and spatial resolutions were used: 1x1° at a 12-hourly temporal resolution, 0.5x0.5° at a 12-hourly temporal resolution, and 1x1° at a 3-hourly temporal resolution. This approach allows us to determine whether an increased spatial and/or temporal resolution is more beneficial for the same increase in observational constraints.

Figure 2a shows the whole-domain FSS time series for the three spatial and temporal resolutions considered in the study. Comparing the FSS for the 0.5x0.5°and 1x1° model runs using 12-hourly time resolution, we find that the model using 0.5x0.5° degree resolution has a higher FSS value than that using 1x1° resolution 85% of the time. Given this difference, the 1x1° degree, 12-hourly resolution will not be further considered in this study. The mean FSS for the model using 0.5x0.5° spatial and 12-hourly temporal resolution is 0.70, compared to 0.81 for the model using 1x1° spatial and 3-hourly temporal resolution. To determine whether the difference in FSS values is statistically significant, we apply a bootstrapping algorithm to compare the difference between the means. The zero mean difference between the mean FSS values lies outside of the 95% confidence interval, thus, the model using 0.5x0.5° spatial and 12-hourly temporal resolution is more skillful at a statistically significant level. As such, we use this resolution for the remainder of the study (see supplemental information, SI, for more details).

To determine the proper kernel size for the U-Net, we calculate the whole-domain FSS for the CPLRSTW model run, using a 1x1, 3x3 and 5x5 kernel, shown in Figure 2b. The 1x1 kernel model is clearly less skillful than the 3x3 and 5x5 kernel models, while the latter two have fairly similar FSS values across the time series. While the 5x5 kernel model run has the highest FSS value at 63% of the time points, compared to the 3x3 kernel model run having the highest FSS value at 37% of the time points, but the mean FSS values are 0.69 and 0.68, respectively. The marginal improvement in skill of the 5x5 kernel model is counteracted by an increase in computational efficiency. The 5x5 kernel on average takes ~2.75 hours to complete a run whereas the 3x3 kernel model takes 1.25 hours to complete a run for the same input variables (CPLRSTW CNN). As a result of the increased runtime for a relatively marginal improvement in skill, we use the 3x3 kernel during this study.

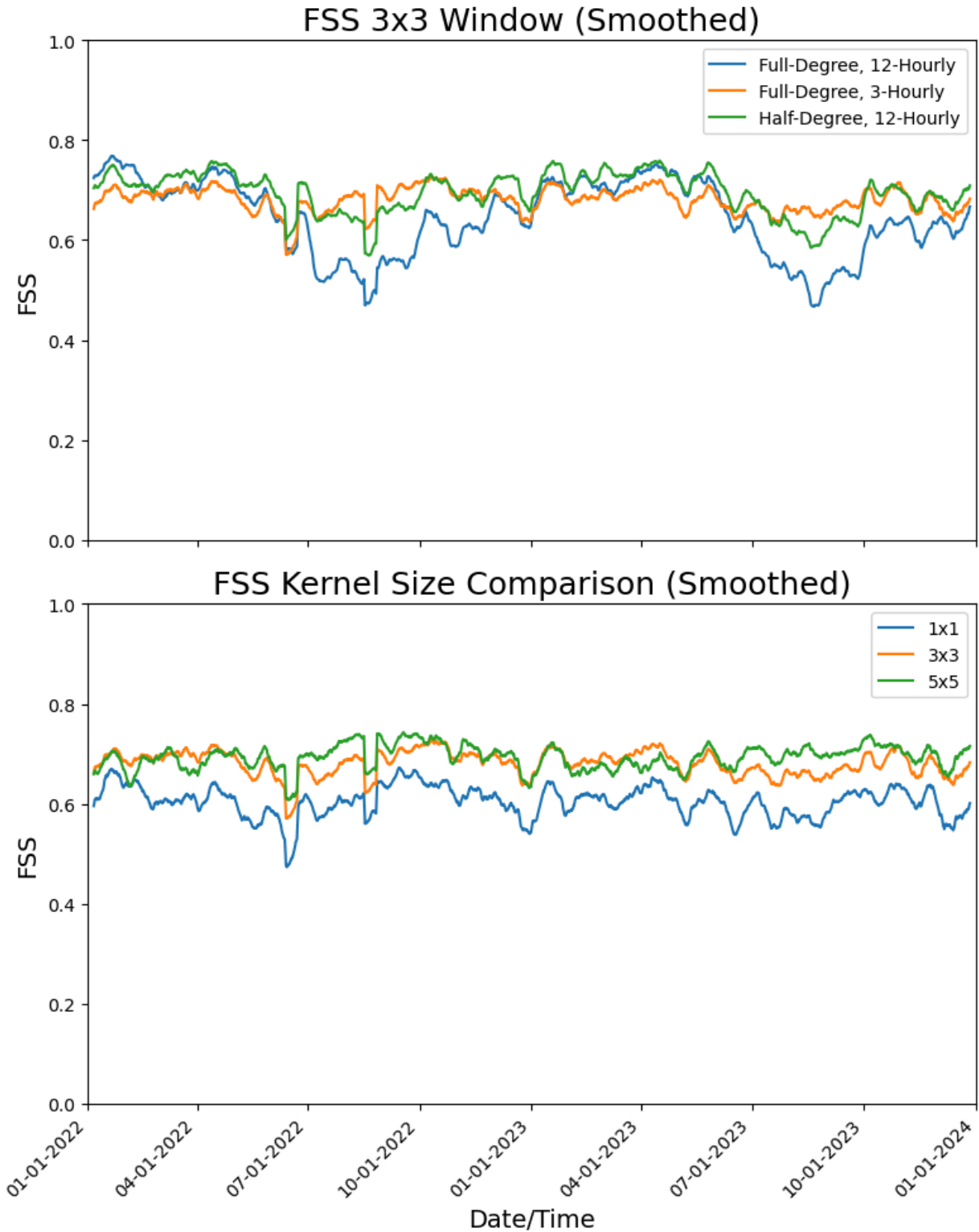

Figure 2: (a) Whole-domain FSS values smoothed with a 10-day running mean for the different spatial and temporal resolutions considered in this study. (b) Whole-domain FSS values of the CPLRSTW CNN model run using different kernel sizes, smoothed with a 10-day running mean.

Using the selected 0.5x0.5° spatial and 12-hourly temporal resolution, and 3x3 kernel, we test three loss functions: mean squared error (MSE), mean absolute error (MAE) and the Dice loss. FSS values for each loss function considered were calculated. The FSS values from the MSE were significantly better than the FSS values from the MAE, while the Dice loss produced lightning stroke density values that were significantly higher than the observed values, resulting in unrealistic lightning stroke density patterns across the entire domain. Thus, we use MSE for all subsequent model runs (see SI).

## RESULTS

To evaluate model performance using subdomains that group similar regions together, we divide the study area into latitude bands: the northern extratropics (20° N to 52° N), tropics (20° S to 20° N) and southern extratropics (52° S to 20° S). We further separate land and oceanic regions in each of these subdomains to further test model performance where different environmental factors have been shown to impact the amount of lightning [*Romps et al.*, 2014]. A grid cell is classified as a land grid cell if the land-sea mask had a value greater than or equal to 0.5 for the pixel, while we classify grid cells with a land-sea mask value less than 0.5 as oceanic.

**Reproducing Lightning Climatology**

Figure 3 shows the 2022-2023 mean lightning stroke density distributions from WWLLN observations, and those predicted by the R14 parameterization and the CP and CPLRSTW CNNs. The colorbars for each plot are log-scaled, and the white values represent zero strokes km$^{-2}$ yr$^{-1}$. The relative distributions of the lightning stroke densities are similar, with some slight over and underestimations throughout the maps in all three model outputs. The R14 parameterization heavily overestimates lightning stroke density over the tropical oceans, with this area representing the largest deviance from the observed lightning stroke density in any model. On average, the R14 parameterization overestimates observed lightning stroke density by $1.4*10^{-4}$ strokes km$^{-2}$ yr$^{-1}$ per pixel, while the CP CNN overestimates by $1.4*10^{-5}$ strokes km$^{-2}$ yr$^{-1}$, representing a factor of 10 larger overestimation by the R14 parameterization compared to the CP CNN. The CPLRSTW CNN slightly underestimates lightning at a rate of $2.2*10^{-6}$ strokes km$^{-2}$ yr$^{-1}$, representing a factor of 6 improvement in the average estimation error over the CP CNN. In spite of these domain mean biases, both CNNs replicate the spatial distribution of lightning stroke density well, with significant improvements over the tropical oceans being visible when compared to the R14 parameterization.

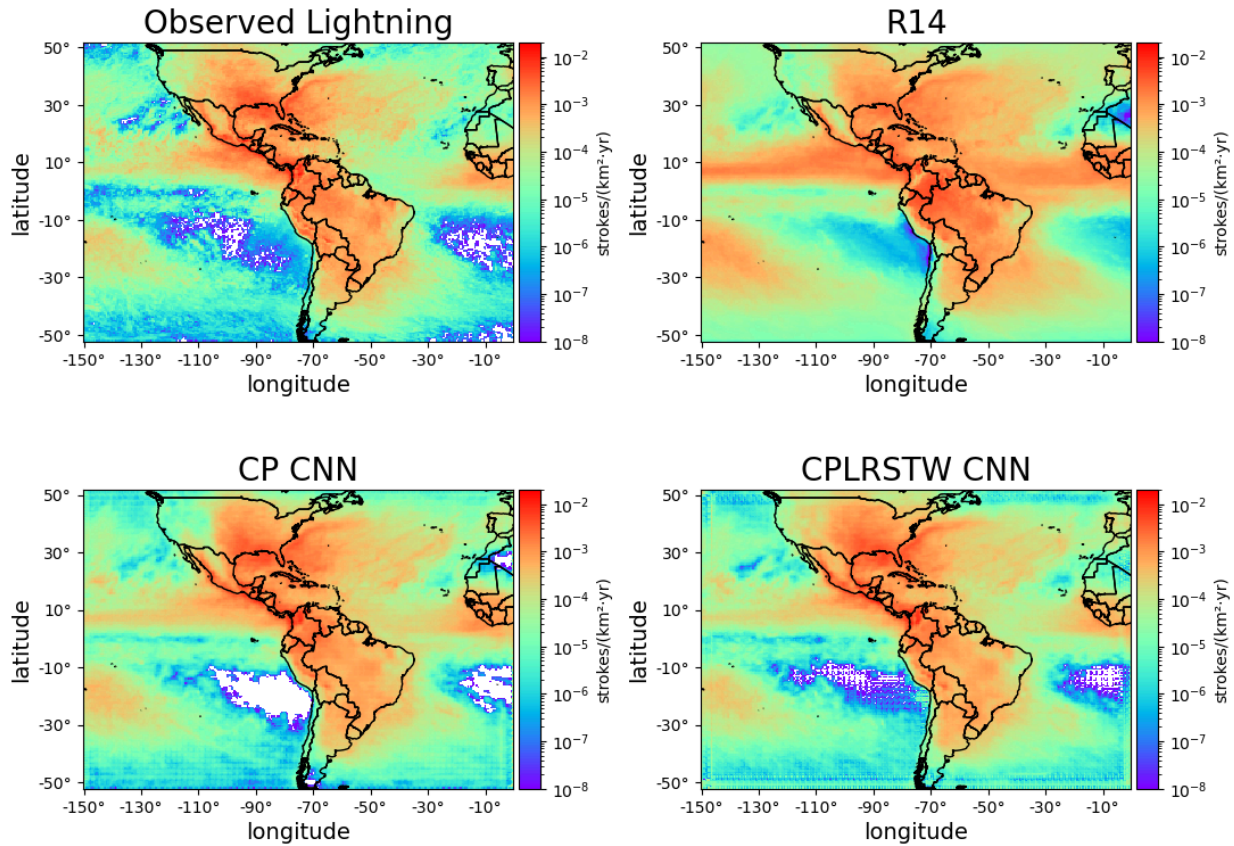

Figure 3: 2022-2023 mean lightning stroke density (strokes km$^{-2}$ yr$^{-1}$) maps from WWLLN (top left), the R14 parameterization (top right), the CP model run of the CNN (bottom left) and the CPLRSTW model run of the CNN (bottom right)

    The model-observation difference maps in Figure 4 more clearly show the regionally varying successes and deficiencies of the different prediction methods. Stippling in Figure 4 represents pixels with statistically insignificant differences between predictions and observations, determined through use of a bootstrapped confidence interval. Over land, stark differences emerge in the success of the R14 parameterization and the CNNs, especially over the Amazon, in which the R14 parameterization heavily overestimates lightning stroke density at a statistically significant level, while most of the statistically significant differences between the CNN predictions and the observations are due to both CNNs underestimating the observations in the region. Interestingly, the R14 parameterization also underestimates lightning stroke density in southern Mexico and near the Gulf Coast of the United States. The CPLRSTW CNN tends to overestimate lightning stroke density over the eastern United States. The western United States has high numbers of statistically significant differences in all models, but the magnitude of these differences is small due to the lower average lightning stroke density in the region compared to other land areas in the domain.

    Major differences exist between the predictions and observations in the northern and tropical Atlantic oceans, where the R14 parameterization routinely overestimates the lightning

stroke density, most heavily near the equator. The distinct value of using a deep learning algorithm is evident in the comparison of the R14 parameterization performance in the tropical oceans to that of the CP and CPLRSTW model runs. Both CNN approaches have fewer pixels with predictions that are statistically significantly different from observations in the tropical Atlantic, while the R14 predictions significantly overestimate observations throughout the tropical Atlantic. There is some overestimation in the Gulf Stream by the CPLRSTW model, but the mean bias decreases further offshore of the U.S. coast over the North Atlantic domain. Using the land-sea mask in the CNN model appears most important for reducing the overestimation over the oceans (see supplemental figures in the SI), as the CPL model significantly underestimates lightning stroke density in the North Atlantic Ocean, whereas all other CNN runs with CP and one additional training variable overestimate lightning in this region.

We calculate the mean bias of statistically significant differences for each prediction method and sub-domain, the results of which can be found in Table 2. In general, as noted above, the R14 parameterization tends to have an order of magnitude larger mean bias (positive) compared to the CNN approaches, which have either positive or negative biases depending on the CNN model and region. Notably, over the northern domain, the average mean bias in the R14 parameterization is three orders of magnitude less than that of both CNN approaches, but this is due to large biases in both directions cancelling each other out.

| Domain | R14 | CP CNN | CPLRSTW CNN |
|---|---|---|---|
| Whole Domain | $1.7*10^{-4}$ | $-3.0*10^{-4}$ | $-3.4*10^{-4}$ |
| Land | $1.4*10^{-4}$ | $-8.3*10^{-5}$ | $-6.1*10^{-5}$ |
| Ocean | $1.8*10^{-4}$ | $7.1*10^{-5}$ | $2.4*10^{-5}$ |
| Northern | $-7.6*10^{-7}$ | $-3.1*10^{-4}$ | $-3.5*10^{-4}$ |
| Tropical | $3.3*10^{-4}$ | $-4.1*10^{-4}$ | $-5.4*10^{-4}$ |
| Southern | $1.0*10^{-4}$ | $-9.4*10^{-5}$ | $-8.4*10^{-5}$ |

Table 2: Mean bias of statistically significant differences between modeled and observed lightning stroke density in units of strokes km$^{-2}$ yr$^{-1}$.

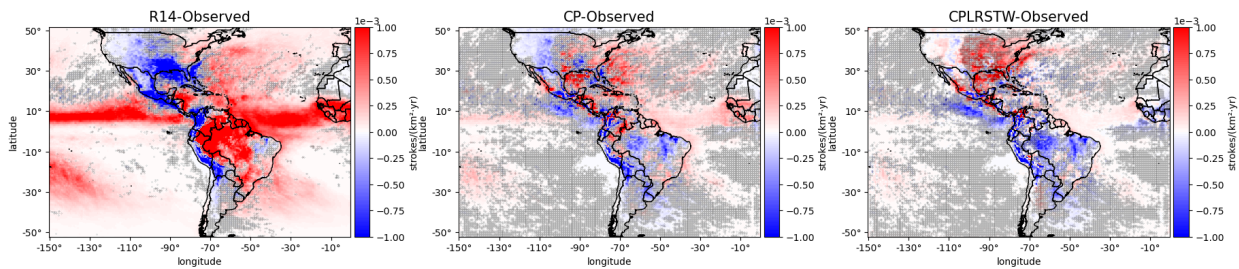

Figure 4: Difference maps showing R14-observed (left), CP CNN-observed (middle) and CPLRSTW CNN-observed (right) mean lightning stroke densities (strokes km$^{-2}$ yr$^{-1}$). Stippling indicates non-significant differences using 95% confidence intervals.

Figure 5 shows heatmaps comparing the observed 12-hourly, 1x1° lightning stroke densities with the model output on a point-by-point basis for the R14 parameterization, the CP model and the CPLRSTW model during the years 2022 and 2023 for the whole domain, and land and ocean domains separately. Each plot contains a 1:1 line with the corresponding r$^2$-value in the legend of the corresponding plot. Consistent with the above statistics, over the whole domain, the R14 parameterization overestimates the lightning stroke density more often than both the CP and CPLRSTW CNN runs, with 84% of the mean lightning stroke density values from the R14 parameterization being higher than the observed lightning stroke densities from WWLLN, compared to 64% and 60% of the modeled lightning stroke densities being higher than the observed lightning stroke densities for the CP and CPLRSTW CNN runs respectively. The shape of the distribution from the R14 parameterization heatmap suggests that a linear relationship may exist between predicted and observed lightning stroke densities in a high lightning regime. However, the R14 parameterization tends to heavily overestimate lightning stroke density in the low lightning regime, leading to the distribution being skewed with a lower r$^2$ value of 0.4. In contrast, both CNN model output lightning stroke densities are more regularly distributed across the full distribution of observed lightning stroke densities, with a slight positive model bias being present throughout the entire distribution and r$^2$ values of >0.85 for both CNNs. Given that the CP CNN is trained on the same two variables used to generate the R14 parameterization, it is clear that the assumption of a linear relationship between the multiplicative product of CAPE and precipitation is not sufficient to describe lightning outside of high lightning regimes. Allowing a non-linear (and spatially varying) relationship between CAPE and precipitation can produce lightning stroke density predictions which rival the accuracy of those from a model trained on a five times larger parameter space (e.g. CPLRSTW).

The performance of the models over land is similar to those from the whole domain, with the CPLRSTW model having the best r$^2$ value, followed by the CP model and the R14 parameterization having the worst r$^2$ value. However, the r$^2$ values are lower over land than over the entire domain, with the R14 parameterization having a r$^2$ of 0.2, the CP model having a value of 0.76, and the CPLRSTW model having an r$^2$ value of 0.8. All distributions are generally centered around the 1:1 line against observed lightning stroke density over land, but the spread is greater, likely due to the more limited domain. Relative to the whole domain, the spread of the model predictions versus observations is more symmetric about the 1:1 line for all models over land, indicating a minimal mean bias for predictions over land.

Over the oceans, it is evident that the R14 parameterization consistently overestimates the lightning stroke density, with 94% of the modeled values exceeding the observed values in the same grid cells. The distribution appears to be centered along a line with a lesser slope, suggesting that a modification to the CAPE*precipitation formula may be more accurate over the ocean. The $r^2$ value for the R14 parameterization over the ocean is 0.44, much higher than the $r^2$ value for the R14 parameterization over land, despite the distribution over the ocean not being centered around the 1:1 line. Similar to the whole domain, the CP and CPLRSTW CNNs also tend to overestimate lightning stroke density, but to a lesser degree, with 71% and 63% of the modeled values being greater than their corresponding observed values respectively. Additionally, the spread of the modeled lightning stroke density is centered near the 1:1 line, with the $r^2$ values for the CP and CPLRSTW CNNs being 0.90 and 0.93 respectively. A possible reason for the improved $r^2$ values over the ocean compared to the land is that there are roughly four times as many oceanic points in the domain as there are land points (49,070 points over the ocean compared to 13,330 points over land), so the underestimations seen by the models weigh more heavily over land than if an underestimation of the same magnitude were to be seen over the ocean.

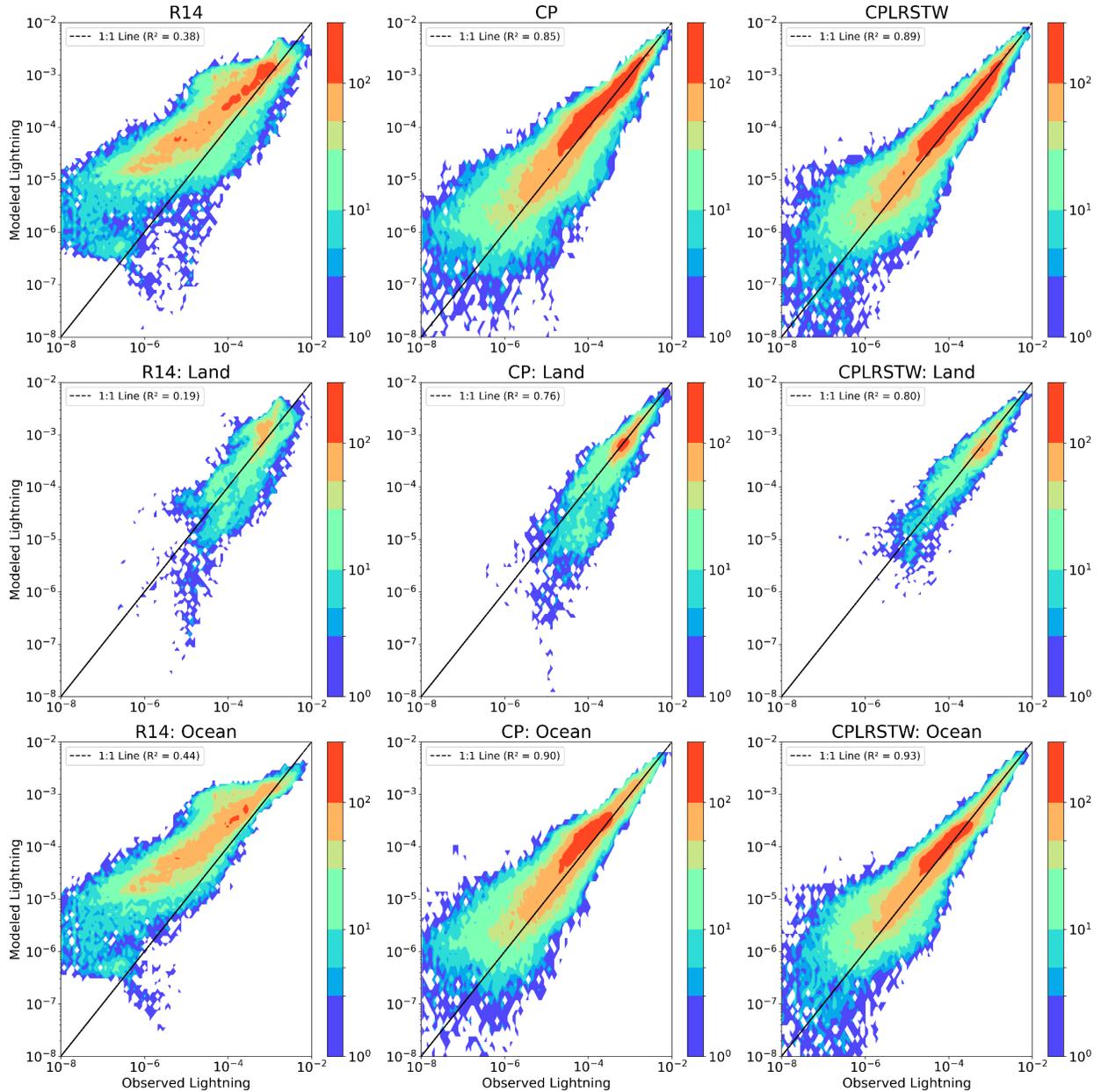

Figure 5: Heatmaps showing observed vs. modeled lightning for the R14 parameterization (left column), the CP model (middle column) and the CPLRSTW model (right column), separated into the whole domain (top row), land (middle row) and ocean (bottom row). The black line depicts the 1:1 line for observed vs. modeled lightning. Axes represent lightning stroke density (strokes km$^{-2}$ yr$^{-1}$) while the color represents the number of point density of observations.

Figure 6a shows the domain-averaged FSS values for the R14 parameterization, CP CNN model output and CPLRSTW model output smoothed over 10 days. Both CNN models show significant improvement over the R14 parameterization, as the R14 parameterization has the lowest FSS value for the entirety of the evaluation period. The CPLRSTW CNN outperforms the

CP CNN, with the highest domain-wide FSS value being seen in the CPLRSTW at 73% of time points, compared to only 27% of times for the CP CNN.

Figure 6b shows the domain-averaged MFSS values, with the error bars indicating the 95% confidence interval of the bootstrapped mean. The lowest bin, representing zeroes in the observed dataset, is captured very well by the CP and CPLRSTW CNNs, with MFSS values of 0.98 and 0.97 respectively, compared to 0.69 for the R14 parameterization. In the range from $10^{-8}$ to $10^{-5}$ strokes $km^{-2}$ $yr^{-1}$, there are no observed lightning stroke density values in the WWLLN dataset. However, the models do predict lightning in this range, and as such the MFSS values are 0 indicating no skill in this low lightning regime. The R14 parameterization contains the largest number of overestimations, likely due to its multiplicative nature opposed to an image-based scheme such as the CNN, with 32.1% of the total observations occurring in this range in the R14 parameterization, compared to only 0.78% and 1.8% for the CP and CPLRSTW CNNs respectively (see supplemental). The R14 parameterization consistently has statistically significant less skill in each bin.

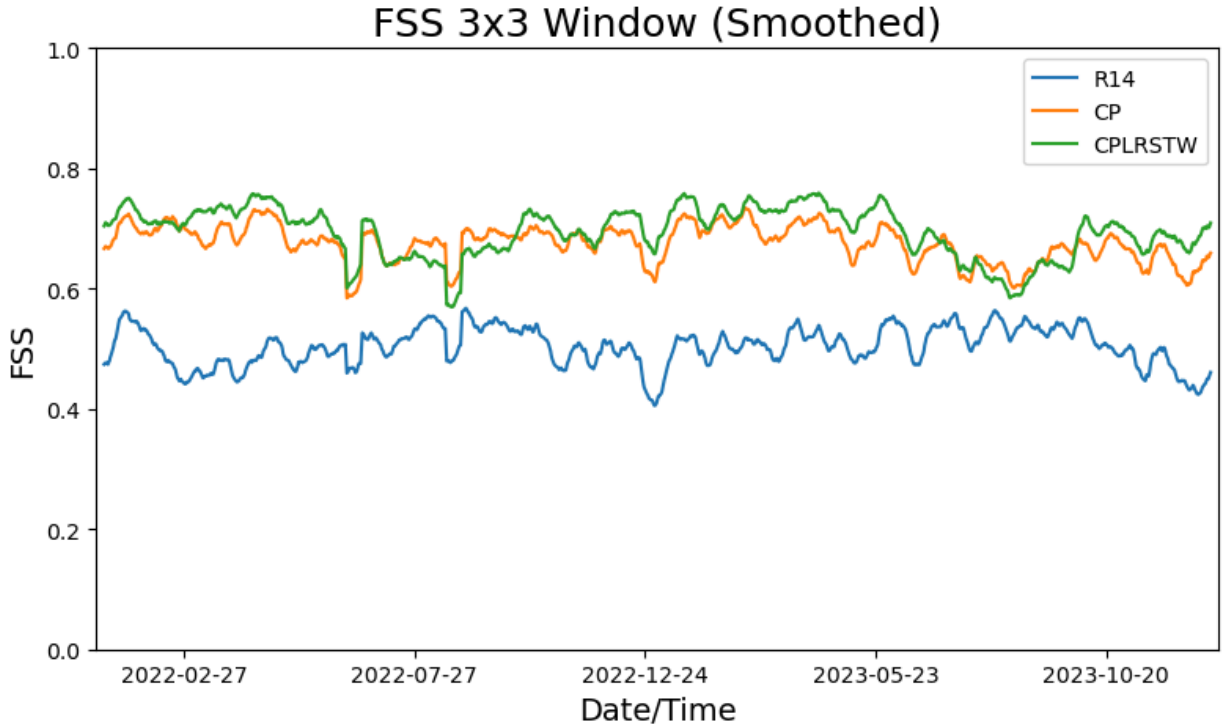
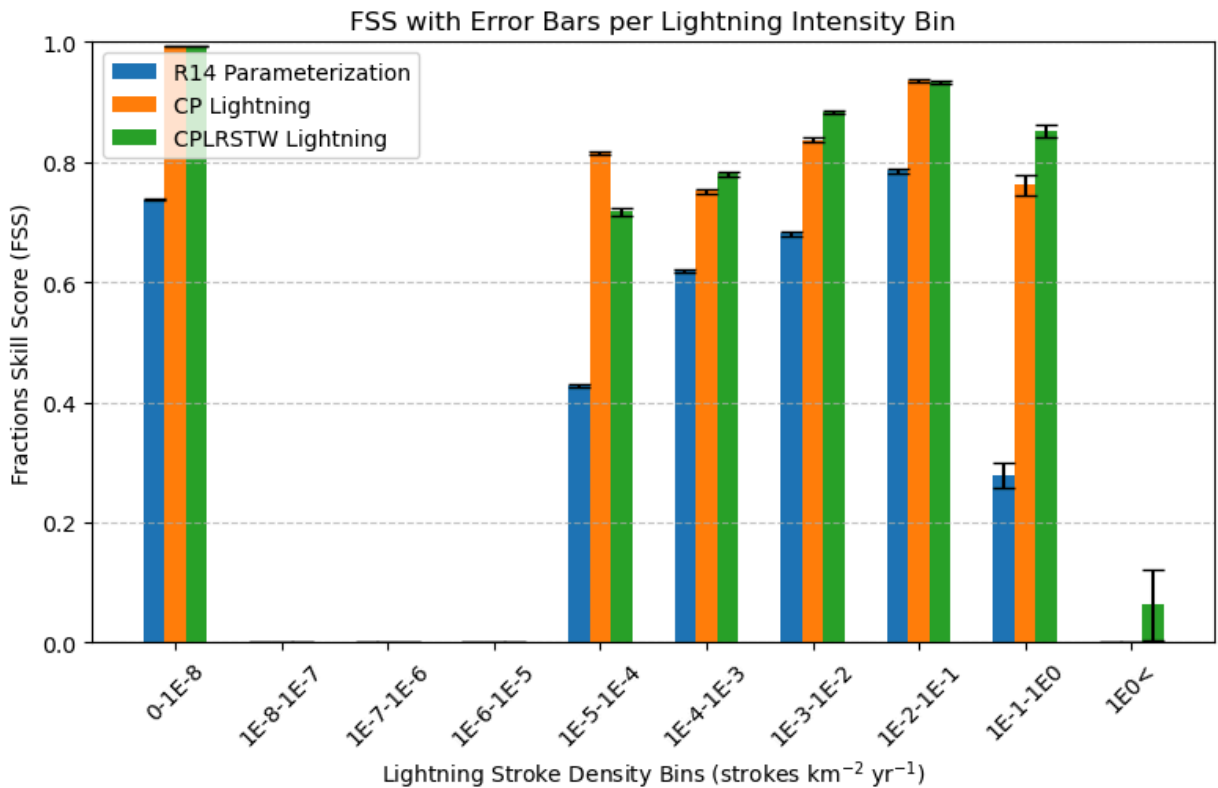

Figure 6: (a) Smoothed domain-wide FSS values for the R14 parameterization (blue), CP model run (orange) and CPLRSTW model run (green) and (b) Whole domain MFSS values with 95% confidence interval as error bars

**Case Studies**

To investigate the ability of the models to capture lightning in temporally and spatially evolving weather events as opposed to the annual climatology or domain-wide time series of model skill, we examine model performance in two separate convective events covering different domains. One event occurs in North America from June 27 to July 1, 2023, and the second occurs in South America from January 1 to January 5, 2023.

A snapshot from the North American event is shown in Figure 7, with the mean lightning observed by WWLLN over the eastern continental United States on June 29$^{th}$, 2023 in the top left, and predictions from the R14 parameterization (top right), the CP CNN model (bottom left), and the CPLRSTW CNN model (bottom right) for comparison. In the observed lightning, there is a hook-like feature extending from the Great Plains into Louisiana and Mississippi, as well as a linear feature off the east coast of the United States extending into Florida. These high lightning areas are qualitatively captured well by all three models, with the spatial accuracy of these being mixed. While there is no observed lightning in the southwestern part of the domain, the CP CNN produces lightning all throughout the domain with the exception of the mid-Atlantic states north through Ontario. Qualitatively, the R14 parameterization appears to be the best performing model, especially south of the hook-like feature where both the R14 model and observations indicated no significant lightning. All models produce a feature in Maine and eastern Canada, which is not present in the observed lightning stroke density field, and fail to capture the sharpness in the lightning gradients between the hook feature in the central United States and the secondary observed feature along the United States-Canada border to the north.

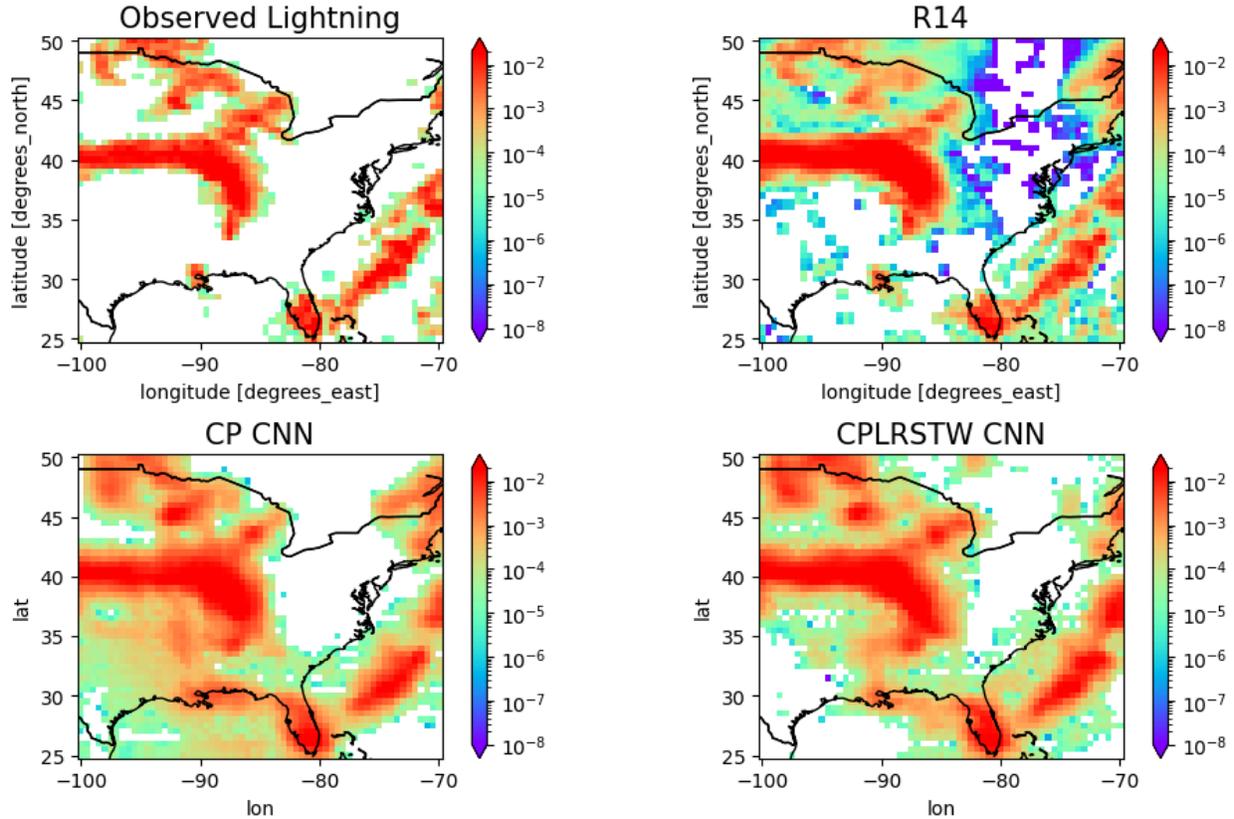

Figure 7: Lightning stroke density (strokes km$^{-2}$ yr$^{-1}$) over eastern North America from WWLLN (top left), the R14 parameterization (top right), the CP CNN (bottom left) and the CPLRSTW CNN (bottom right) on June 29th, 2023

Figure 8a shows the FSS values for the North American event. The R14 is the best performing model in this event, as it has the highest FSS values at 9 of the 10 time points sampled in the event, and the CPLRSTW CNN outperforms the CP CNN in 9 of the 10 time points in the event. The strength of the R14 parameterization is evident in this event, given its simplicity and performance over the continental United States.

Figure 8b shows the MFSS for the North American event, which has a mix of results. The lower and higher lightning regimes do not show a best model at a statistically significant level, with all three models exhibiting overlapping error bars in the $10^{-2}$ to $10^{-1}$ strokes km$^{-2}$ yr$^{-1}$ bin as well as the $10^{-1}$ and $10^0$ bins, while the 0 to $10^{-8}$ bin shows virtually no difference between the R14 parameterization and the CPLRSTW CNN, both of which perform better than the CP CNN. The R14 parameterization is the best performing model in the bins ranging from $10^{-5}$ to $10^{-2}$ at a statistically significant level, with the CPLRSTW being the second-best model at a statistically significant level in the bin from $10^{-4}$ to $10^{-3}$. These results again show the strength of the R14 parameterization over the continental United States, which this North American event heavily features.

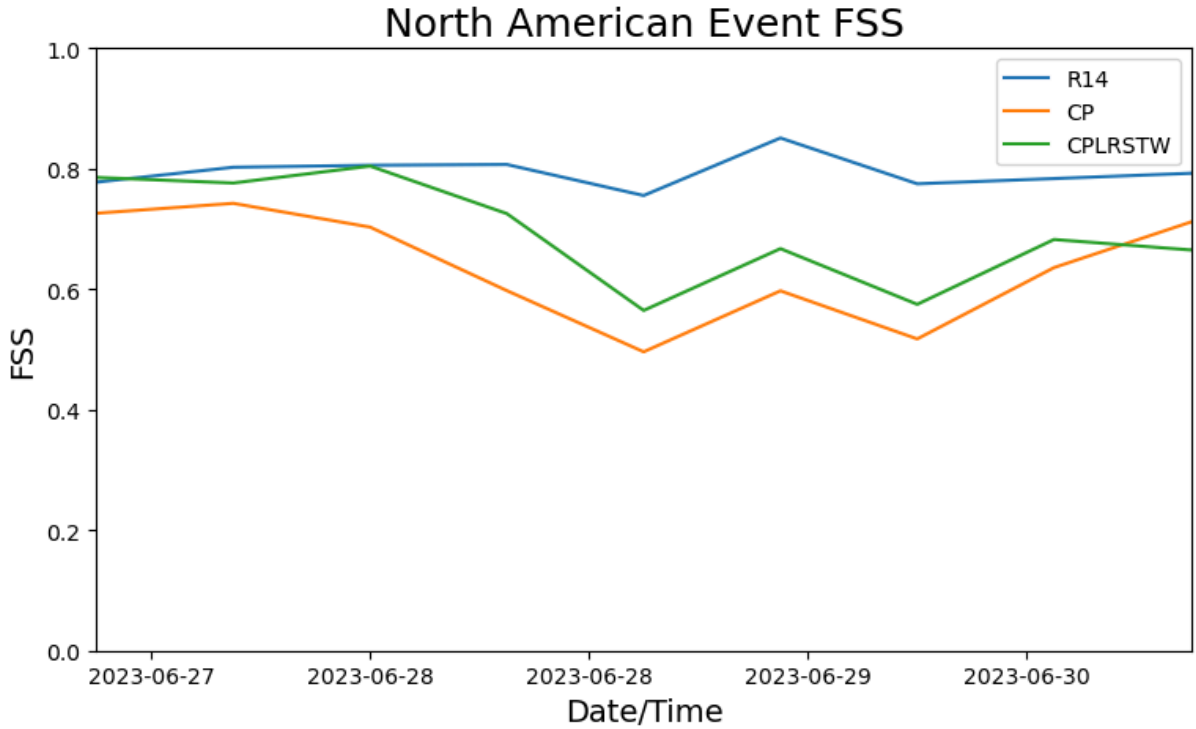

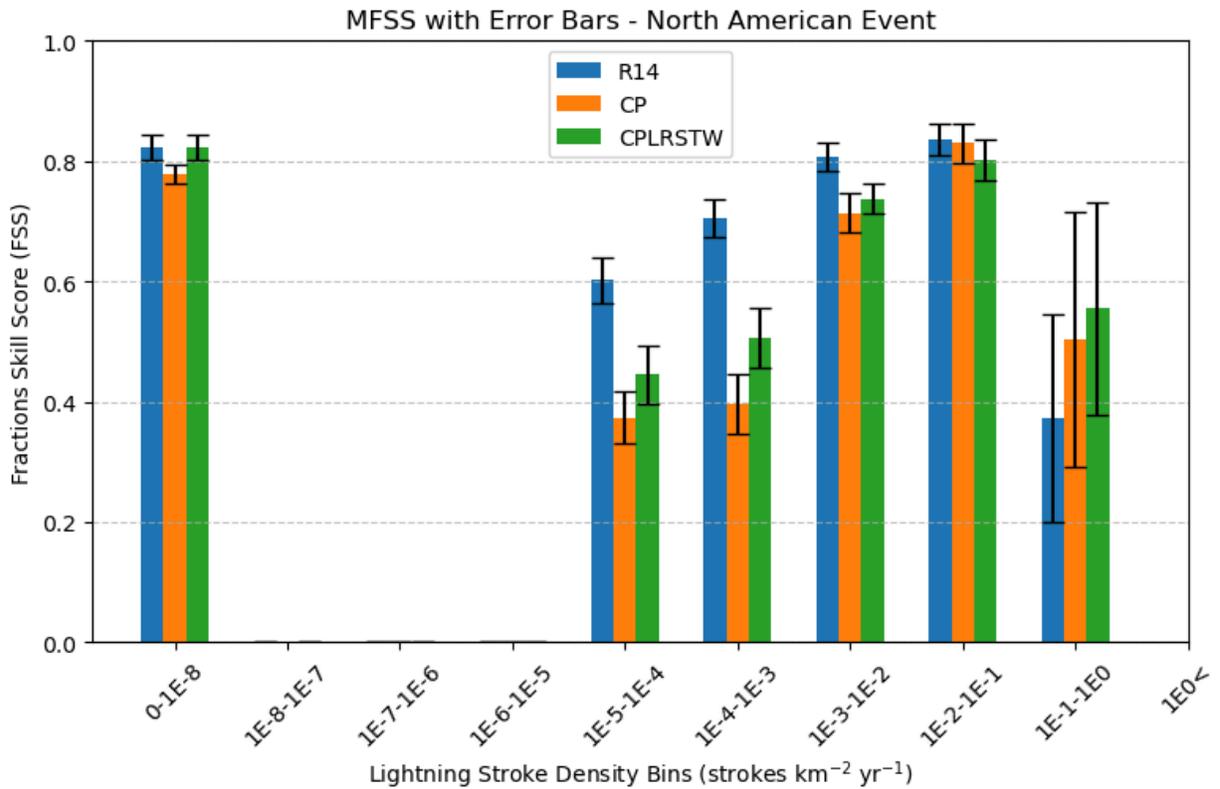

Figure 8: (a) FSS values for the North American event, from June 27th - July 1st, 2023 and (b) MFSS values for the North American event

The selected South American lightning event helps illustrate additional complexities in capturing lightning stroke densities in the tropical and southern land regions. Figure 9 shows lightning observations and predictions in northern South America from January 5th, 2023. North of 20° S, the event brings high lightning stroke densities over land throughout the western coast of South America and Brazil. There are sharp gradients in the lightning stroke density over Brazil, which are not replicated by the R14 parameterization or the CNN models. There is also a high lightning feature during this event over the ocean north of the Brazilian coast. In the oceanic areas during this event, the R14 parameterization struggles heavily, producing lightning in the Atlantic Ocean in regions where there is no observed lightning, as well as overproducing lightning off of the coast of Brazil where lightning is observed. While the feature is captured by both the CP and CPLRSTW models, the observed lightning stroke density in this oceanic feature is overestimated by both the CP and CPLRSTW CNN, albeit to a greater degree by the CP CNN. This feature extends into the Guianas as well, where the magnitude of the overestimation by the CP CNN is even larger than over the ocean.

A high lightning feature is also observed during this event along the border of Chile and Argentina, which is not replicated by the CP model, but is produced by the R14 parameterization and the CPLRSTW model. This feature in the observed lightning is rather narrow, consisting of only a couple of pixels in longitude. As a result, while the feature appears in the CPLRSTW CNN, the longitudinal extent of the feature is larger than it should be, consistent with the known tendency of CNN models to lose sharpness. The R14 parameterization produces lightning where this feature occurs, but not to the correct magnitude, as well as over a broader region than the observed feature. The inclusion of the land-sea mask and relative humidity during the training of the CPLRSTW CNN both result in a similar lightning feature, while windshear, two-meter temperature and warm cloud depth as variables all produce less accurate lightning features than the CPLRSTW CNN.

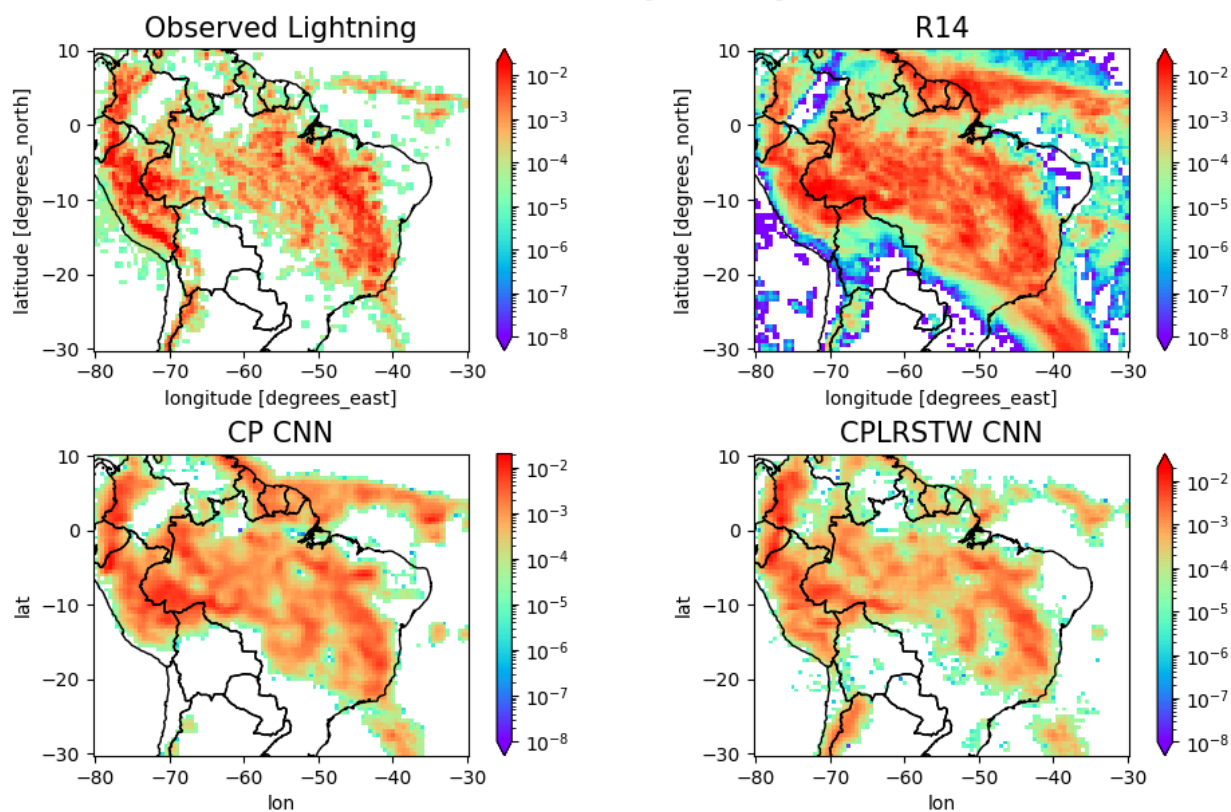

Figure 9: Lightning stroke density (strokes km$^{-2}$ yr$^{-1}$) over northern South America from WWLLN (top left), R14 (top right), the CP CNN (bottom left) and the CPLRSTW CNN (bottom right) on January 5th, 2023

Figure 10a shows the FSS values for the South American event. The CPLRSTW CNN has the best FSS values throughout the event, followed by the CP CNN and finally the R14 parameterization, but this diverges from the lightning stroke density maps in Figure 9, in which the R14 parameterization is shown to overestimate many areas in the low lightning regime, especially over the oceans and around the edges of the main lightning feature which stretched across Brazil. The similarities between the FSS values in this event shows the need for the MFSS adjustment which accounts for the proximity to the correct value, because areas of persistent overestimation by the models are not penalized to an adequate degree in the FSS metric.

Figure 10b shows the MFSS for the South American event, with more distinct differences between the skills of the models. In the low lightning regime, the R14 parameterization is shown to be significantly worse than both the CP and CPLRSTW CNNs, with a MFSS value of only 0.57 in the 0 to $10^{-8}$ bin, compared to values of 0.92 and 0.94 for the CP and CPLRSTW CNNs respectively. The CPLRSTW CNN has the best MFSS value at a statistically significant level in the 0 to $10^{-8}$, $10^{-5}$ to $10^{-4}$, and $10^{-3}$ to $10^{-2}$ bins, while all other bins show no best model at a statistically significant level. In the $10^{-4}$ to $10^{-3}$ bin, the CPLRSTW CNN is significantly better at capturing lightning stroke density than the R14 parameterization, but not the CP CNN. These results show that the R14 parameterization does not capture lightning stroke density patterns over the tropical oceans and in South America as well as the CNN models, with a large number of low lightning stroke density values being overestimated by the R14 parameterization and thus leading to the lower FSS and MFSS scores. As noted above, the R14 parameterization had the highest mean FSS and MFSS scores for the North American lightning event, but it exhibits the lowest skill of the models for the South American lightning event. As of yet, we do not have a testable explanation for this difference in R14 performance over land in North America compared to South America. One possible explanation is that the CNN models are able to partially compensate for errors in the reanalysis products, such as CAPE, that might stem from regions with fewer observational constraints. Alternatively, the dynamical relationship of CAPE to storm updraft intensity and electrification processes may be fundamentally different over the Amazon than over the North American land masses.

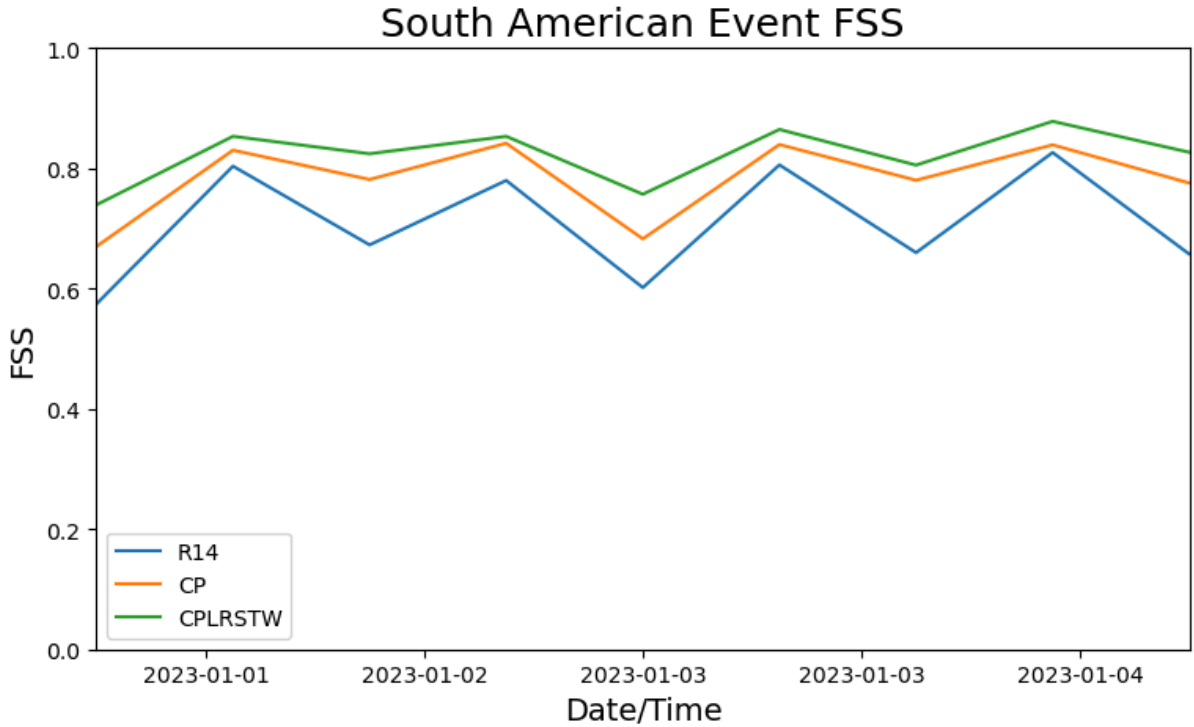

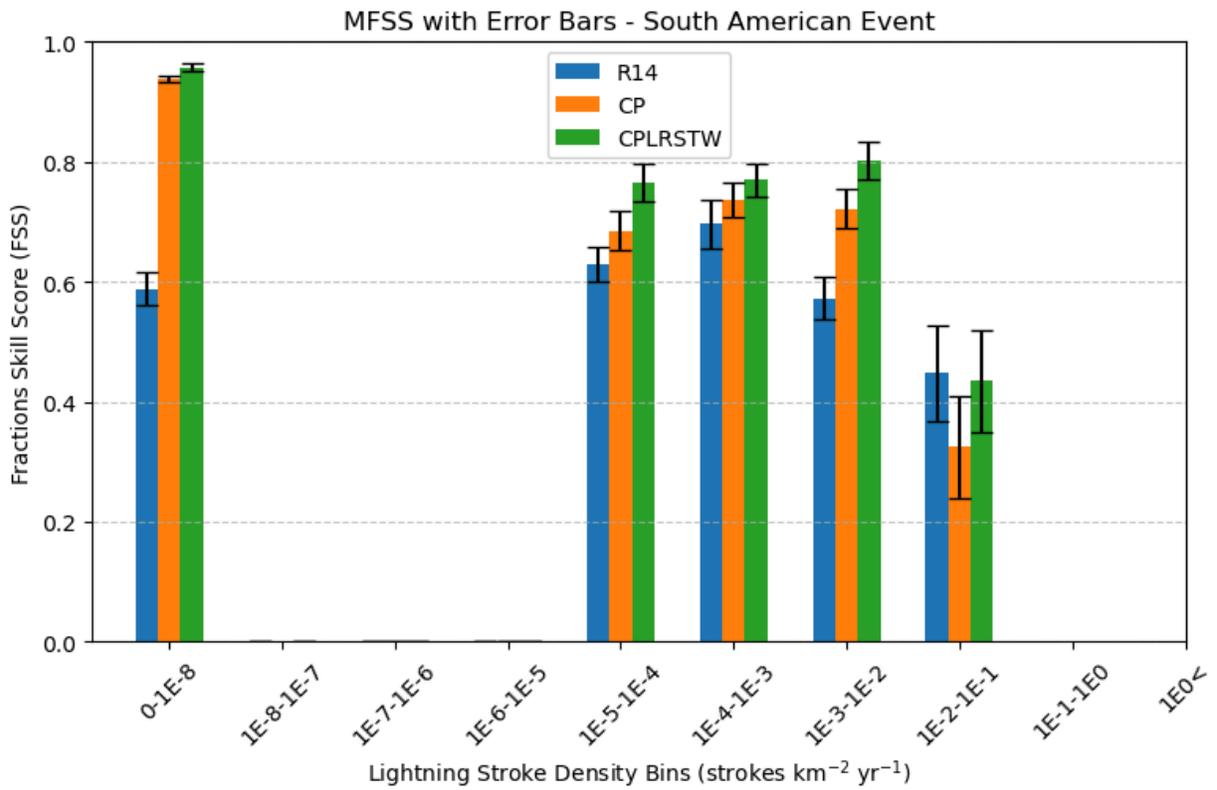

Figure 10: (a) FSS values for the South American event, from January 1st, 2023 to January 5th, 2023 and (b) MFSS values for the South American event

# CONCLUSIONS

We developed a pair of CNNs using meteorological input variables which capture lightning events with high skill at a 0.5x0.5° spatial and 12-hourly temporal resolution. The CNNs used in this study provide general improvement over the R14 parameterization, with significant improvement seen over the oceans and South America especially. The R14 parameterization is found to perform well over the continental United States, but changes to this parameterization are needed when moving to other domains. The difference in skill between the two CNN models (CP vs. CPLRSTW) is marginal in most cases and provides insight into whether other meteorological variables are necessary to accurately capture the lightning patterns. The CPLRSTW CNN tends to outperform the CP CNN but requires the inclusion of five additional variables which requires more disk space. As such, the CP CNN may be recommended if storage is limited and the general pattern of lightning is more important than specificity, such as for global scale high temporal frequency predictions.

As expected, neither of the CNNs used here are able to capture sharp spatial gradients evident in observed lightning patterns. Data used in the CNNs are from reanalysis products with the exception of precipitation, and these datasets typically have smooth fields with regards to these variables. The CNNs also involve a coarsening of the resolution of the data fields during max-pooling, in which four cells are combined into one, which is repeated for the second convolution. As such, smoothing takes place inherently due to the contributions of the input variables in the finer grid cells to these coarser grid cells. Future work may explore alternative neural network architectures to better capture these finer scale features of lightning stroke density.

# ACKNOWLEDGEMENTS

We wish to thank the World Wide Lightning Location Network (http://wwlln.net), a collaboration among over 50 universities and institutions, for providing the lightning location data used in this paper. This work was supported by the NASA Future Investigators in NASA Earth and Space Science and Technology (FINESST) fellowship program under award number 80NSSC23K1550. This work was supported by the National Science Foundation (NSF) under grant number AGS-2113494.

**Supplemental Information**

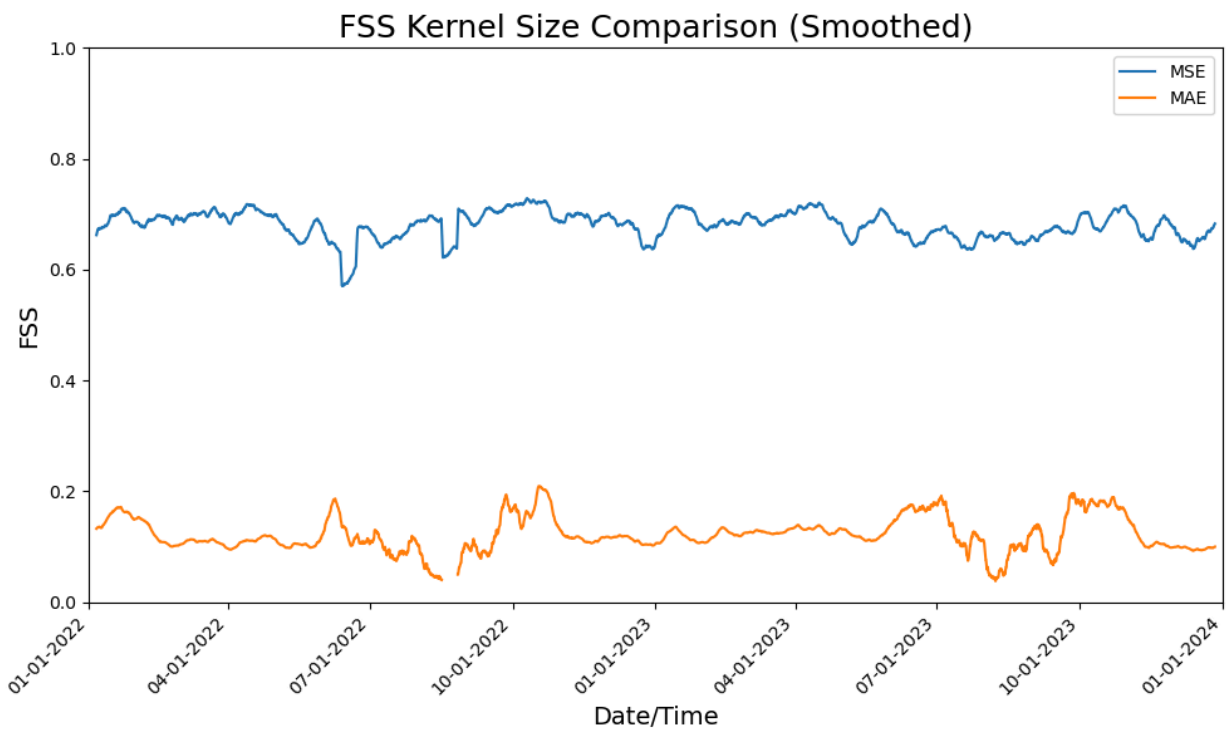

Figure S1: FSS values smoothed over 10 days comparing the MSE and MAE loss functions for the CPLRSTW CNN

Figure S2 shows the bootstrapped distribution of mean difference FSS values between the 1° spatial and 3-hourly temporal resolution and 0.5° spatial and 12-hourly temporal resolution, with the green dashed lines representing the 95% confidence interval and the red dashed line representing zero mean difference between the two models. The mean FSS for the 0.5° spatial and 12-hourly temporal resolution is 0.694, compared to 0.710 for the 1° spatial and 3-hourly temporal resolution. To determine whether the 1° spatial and 3-hourly temporal resolution is better at a statistically significant level, a bootstrapping algorithm is applied to compare the difference between the means. Since the zero mean difference lies outside of the 95% confidence interval, the 1° spatial and 3-hourly temporal resolution model is found to be better at a statistically significant level.

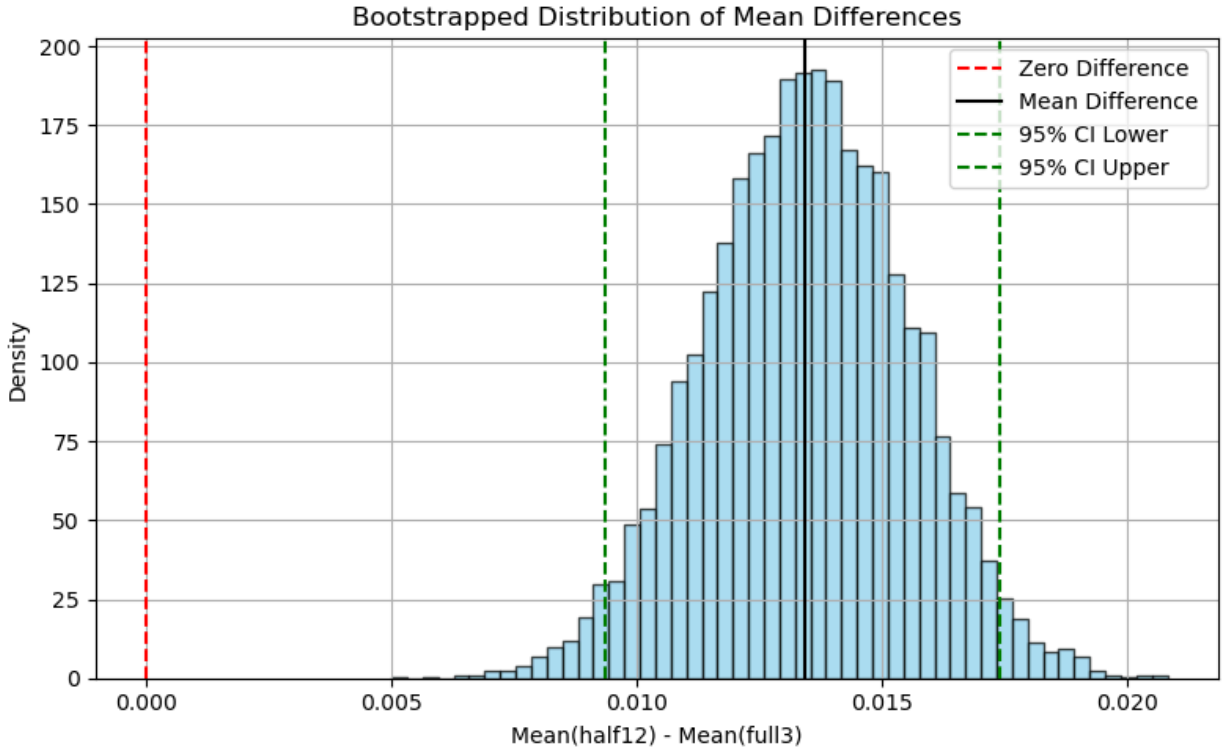

Figure S2: Bootstrapped distribution of mean difference FSS values between the 1° spatial and 3-hourly temporal resolution and 0.5° spatial and 12-hourly temporal resolution

# Lightning Climatology and Difference Maps
## 2022-2023

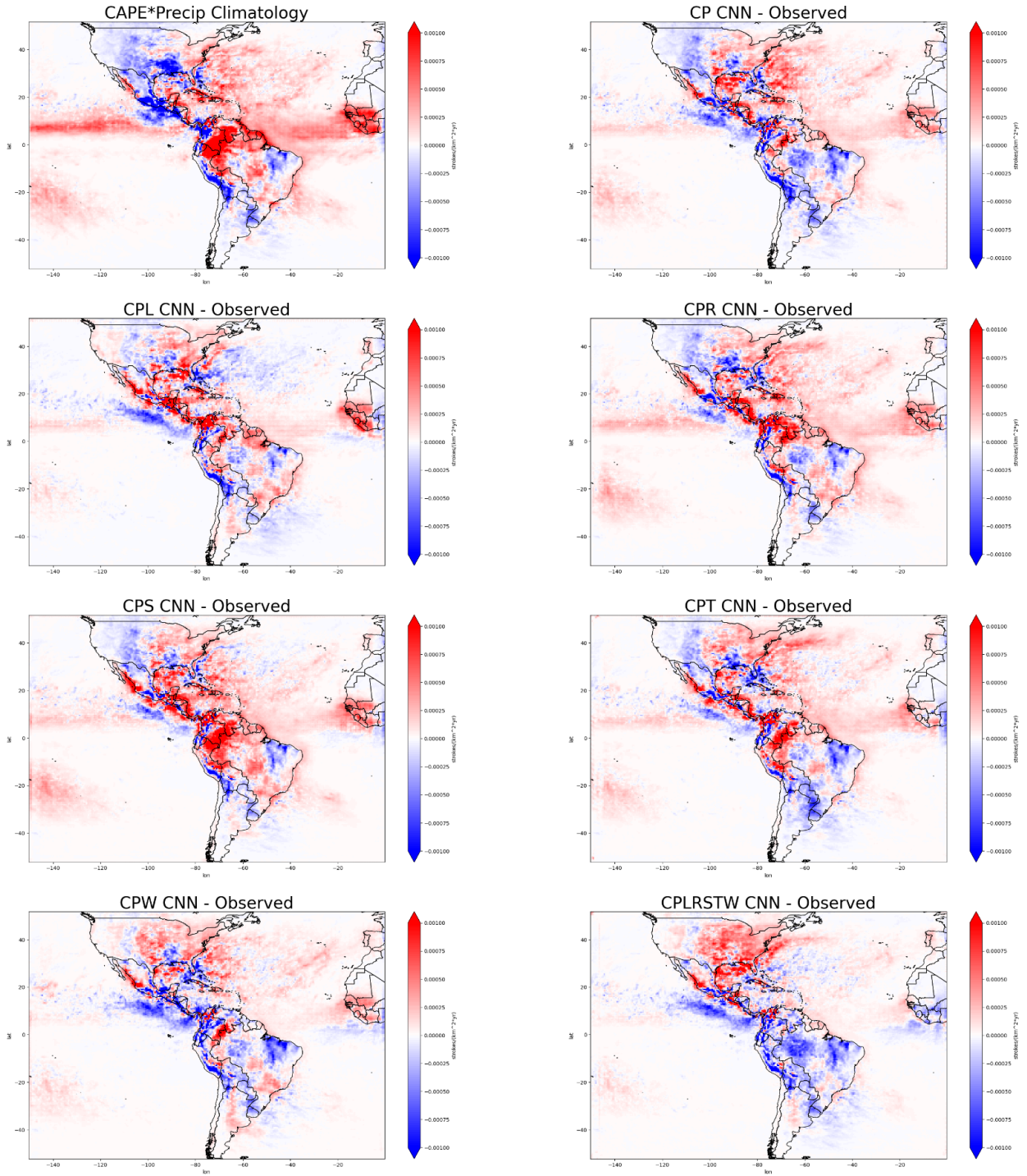

Figure S3: Difference maps showing modeled-observed mean lightning stroke densities (strokes km$^{-2}$ yr$^{-1}$).

|  | Observed | R14 | CP | CPLRSTW |
|---|---|---|---|---|
| $0\text{-}10^{-8}$ | 83,285,559 (91.42%) | 38,249,343 (41.98%) | 77,416,705 (84.98%) | 75,375,891 (82.74%) |
| $10^{-8}\text{-}10^{-7}$ | 0 (0%) | 7,334,534 (8.05%) | 7,115 (0.01%) | 18,333 (0.02%) |
| $10^{-7}\text{-}10^{-6}$ | 0 (0%) | 10,460,645 (11.48%) | 68,618 (0.08%) | 178,359 (0.20%) |
| $10^{-6}\text{-}10^{-5}$ | 0 (0%) | 11,430,296 (12.55%) | 627,679 (0.69%) | 1,438,299 (1.58%) |
| $10^{-5}\text{-}10^{-4}$ | 2,581,685 (2.83%) | 10,071,435 (11.05%) | 3,499,091 (3.84%) | 5,095,018 (5.59%) |
| $10^{-4}\text{-}10^{-3}$ | 2,661,265 (2.92%) | 7,983,288 (8.76%) | 5,514,447 (6.05%) | 5,450,270 (5.98%) |
| $10^{-3}\text{-}10^{-2}$ | 1,994,697 (2.19%) | 4,714,582 (5.17%) | 3,446,724 (3.78%) | 3,044,317 (3.34%) |
| $10^{-2}\text{-}10^{-1}$ | 558,921 (0.61%) | 853,622 (0.94%) | 509,996 (0.56%) | 489,294 (0.54%) |
| $10^{-1}\text{-}10^{0}$ | 21,862 (0.02%) | 6,255 (0.01%) | 13,625 (0.01%) | 14,217 (0.02%) |
| $10^{0}<$ | 11 (<0.01%) | 0 (0%) | 0 (0%) | 2 (<0.01%) |

Table S1: Number of data points in each lightning stroke density bin from WWLLN, the R14 parameterization and the CNN runs, with the percentage of the total number of data points in parenthesis